\documentclass[draftclsnofoot,twoside,onecolumn,letter,12pt]{IEEEtran}
\usepackage{graphicx}
\usepackage{amssymb}
\usepackage{amsmath,cases}
\usepackage{cite}
\usepackage{stfloats}
\usepackage{subfigure}
\usepackage{epstopdf}
\usepackage{mdwlist}
\usepackage{threeparttable}
\usepackage[all,2cell]{xy} \UseAllTwocells
\usepackage{booktabs}
\usepackage{fancyvrb}
\usepackage{rotating}
\usepackage{rotfloat}
\usepackage{pifont}
\usepackage{verbatim}
\usepackage{balance}
\usepackage{bm}

\usepackage{color}
\usepackage{xcolor}
\usepackage{bbm}
\usepackage{dsfont}
\usepackage[mathscr]{euscript}
\usepackage{braket}
\usepackage{mathtools}

\DeclarePairedDelimiter\floor{\lfloor}{\rfloor}

\newcommand{\PeSL}{P_{\left.\mathrm{b}\right|{\mathcal{S}_{i}}}}

\DeclareMathAlphabet{\mathsfbr}{OT1}{cmss}{m}{n}
\SetMathAlphabet{\mathsfbr}{bold}{OT1}{cmss}{bx}{n}
\DeclareRobustCommand{\msf}[1]{%
  \ifcat\noexpand#1\relax\msfgreek{#1}\else\mathsfbr{#1}\fi
}

\makeatletter
\newcommand{\msfgreek}[1]{\csname s\expandafter\@gobble\string#1\endcsname}
\makeatother
\DeclareFontEncoding{LGR}{}{} 
\DeclareSymbolFont{sfgreek}{LGR}{cmss}{m}{n}
\SetSymbolFont{sfgreek}{bold}{LGR}{cmss}{bx}{n}
\DeclareMathSymbol{\salpha}{\mathord}{sfgreek}{`a}
\DeclareMathSymbol{\sbeta}{\mathord}{sfgreek}{`b}
\DeclareMathSymbol{\sgamma}{\mathord}{sfgreek}{`g}
\DeclareMathSymbol{\sdelta}{\mathord}{sfgreek}{`d}
\DeclareMathSymbol{\sepsilon}{\mathord}{sfgreek}{`e}
\DeclareMathSymbol{\szeta}{\mathord}{sfgreek}{`z}
\DeclareMathSymbol{\seta}{\mathord}{sfgreek}{`h}
\DeclareMathSymbol{\stheta}{\mathord}{sfgreek}{`j}
\DeclareMathSymbol{\siota}{\mathord}{sfgreek}{`i}
\DeclareMathSymbol{\skappa}{\mathord}{sfgreek}{`k}
\DeclareMathSymbol{\slambda}{\mathord}{sfgreek}{`l}
\DeclareMathSymbol{\smu}{\mathord}{sfgreek}{`m}
\DeclareMathSymbol{\snu}{\mathord}{sfgreek}{`n}
\DeclareMathSymbol{\sxi}{\mathord}{sfgreek}{`x}
\DeclareMathSymbol{\somicron}{\mathord}{sfgreek}{`o}
\DeclareMathSymbol{\spi}{\mathord}{sfgreek}{`p}
\DeclareMathSymbol{\srho}{\mathord}{sfgreek}{`r}
\DeclareMathSymbol{\ssigma}{\mathord}{sfgreek}{`s}
\DeclareMathSymbol{\stau}{\mathord}{sfgreek}{`t}
\DeclareMathSymbol{\supsilon}{\mathord}{sfgreek}{`u}
\DeclareMathSymbol{\sphi}{\mathord}{sfgreek}{`f}
\DeclareMathSymbol{\schi}{\mathord}{sfgreek}{`q}
\DeclareMathSymbol{\spsi}{\mathord}{sfgreek}{`y}
\DeclareMathSymbol{\somega}{\mathord}{sfgreek}{`w}

\DeclareMathSymbol{\svarsigma}{\mathord}{sfgreek}{`c}

\DeclareMathSymbol{\sGamma}{\mathalpha}{sfgreek}{`G}
\DeclareMathSymbol{\sDelta}{\mathalpha}{sfgreek}{`D}
\DeclareMathSymbol{\sTheta}{\mathalpha}{sfgreek}{`J}
\DeclareMathSymbol{\sLambda}{\mathalpha}{sfgreek}{`L}
\DeclareMathSymbol{\sXi}{\mathalpha}{sfgreek}{`X}
\DeclareMathSymbol{\sPi}{\mathalpha}{sfgreek}{`P}
\DeclareMathSymbol{\sSigma}{\mathalpha}{sfgreek}{`S}
\DeclareMathSymbol{\sUpsilon}{\mathalpha}{sfgreek}{`U}
\DeclareMathSymbol{\sPhi}{\mathalpha}{sfgreek}{`F}
\DeclareMathSymbol{\sPsi}{\mathalpha}{sfgreek}{`Y}
\DeclareMathSymbol{\sOmega}{\mathalpha}{sfgreek}{`W}

\DeclareRobustCommand{\mcal}[1]{%
  \ifcat\noexpand#1\relax\mathnormal{#1}\else\cal{#1}\fi
}
\DeclareRobustCommand{\BM}[1]{%
  \ifcat\noexpand#1\relax\bm{\boldUppercaseItalicGreek{#1}}\else\bm{#1}\fi
}
\makeatletter
\newcommand{\boldUppercaseItalicGreek}[1]{\csname var\expandafter\@gobble\string#1\endcsname}
\makeatother

\newcommand{\V}[1]{\bm{#1}}

\DeclareMathAlphabet{\mathpzc}{OT1}{pzc}{m}{it}

\newcommand{\pDefine}[1]{\mathpzc{#1}}

\newcommand{\pM}{\pDefine{m}}
\newcommand{\pN}{\pDefine{n}}
\newcommand{\pP}{\pDefine{p}}
\newcommand{\pQ}{\pDefine{q}}

\newcommand{\FoxH}[7]{H^{#1,#2}_{#3,#4}
                        \left[
                            {#5}
                            \left|
                            \begin{array}{c}
                                {#6}
                                \\
                                {#7}
                            \end{array}
                            \right.
                        \right]}

\DeclareMathAlphabet{\mathpzc}{OT1}{pzc}{m}{it}

\DeclareMathAlphabet{\mathitsf}{OML}{cmbr}{m}{it}

\definecolor{CCTLABgreen}{RGB}{0,128,0}

\AtBeginDocument{}

\AtBeginDocument{}

\newtheorem{theorem}{Theorem}

\newtheorem{remark}{Remark}

\newtheorem{keynote}{Keynote}

\newtheorem{problem}{Problem}

\DeclareMathOperator{\prob}{\mathds{P}}

\newcommand{\usertext}[1]{\mathrm{#1}}

\newcommand{\R}{\mathbbmss{R}}

\newcommand{\B}[1]{\mathbf{#1}}

\newcommand{\Prob}[1]{\prob\left\{{#1}\right\}}

\newcommand{\Binom}[2]{\usertext{Binom}\left({#1},{#2}\right)}

\newcommand{\PV}[1]{\usertext{Poisson}\left(#1\right)}

\newcommand{\sNorm}[1]{\left|{#1}\right|}

\newcommand{\argmin}{\mathop{\mathrm{arg\,min}}}
\newcommand{\argmax}{\mathop{\mathrm{arg\,max}}}

\newcommand{\bTRe}{\begin{dingautolist}{182}}
\newcommand{\eTRe}{\end{dingautolist}}

\newcommand{\GF}[1]{\Gamma\left(#1\right)}

\newcommand{\Tb}{T_\mathrm{b}}

\newcommand{\Pe}{P_{\mathrm{b}}}
\newcommand{\RIBF}[3]{I_{#1}\left({#2},{#3}\right)}

\begin{document}

\title{
Molecular Communication with Passive Receivers in Anomalous Diffusion Channels
\\[0.5cm]
}
\author{
        Dung Phuong Trinh, 
        Youngmin~Jeong, 
        and 
        Sang-Hyo Kim 
\thanks{
        D.~P.~Trinh and S.~Kim are with the Department of Electrical and Computer Engineering,
        Sungkyunkwan University,
        300, Cheoncheon-dong, Jangan-gu,
        Suwon-si, Gyeonggi-do, 16419 Korea
        (e-mail: \{dtrinh, iamshkim\}@skku.edu).
}
\thanks{
        Y.~Jeong is with the Networks Business, Samsung Electronics Co., Ltd, 
        Yeongtong-gu, Gyeonggi-do, Suwon-si, 16677, Korea
        (e-mail: ymn.jeong@samsung.com).
}

}

\maketitle 

\begin{abstract}
We consider anomalous diffusion for molecular communication with a passive receiver. We first consider the probability density function of molecules’ location at a given time in a space of arbitrary dimension. The expected number of observed molecules inside a receptor space of the receiver at certain time is derived taking into account the life expectancy of the molecules. In addition, an implicit solution for the time that maximizes the expected number of observed molecules is obtained in terms of  Fox's $H$-function.
The closed-form expressions for the bit error rate of a single-bit interval transmission and a multi-bit interval transmission are derived. It is shown that lifetime limited molecules can reduce the inter-symbol interference while also enhancing the reliability of MC systems at a suitable observation time.

\end{abstract}

\begin{IEEEkeywords}
Anomalous diffusion, 
bit error rate, 
Fox's $H$-function, 
inter-symbol interference, 
lifetime molecule, 
molecular communication, 
receptor space.
\end{IEEEkeywords}

\newpage

\section{Introduction}		\label{sec:1}
Molecular communication (MC) is a new communication paradigm that uses tiny-machines which are nanometers to micrometers in size as a transmitter, a receiver, and molecules as a communication carrier \cite{NEH:13:Book}.
Brownian motion (normal diffusion) which describes the free movement of molecules in a fluid medium is usually well adopted as an ideal diffusion scheme in MC. However, we may encounter extraordinary diffusion or anomalous diffusion, which does not obey normal diffusion principles, in crowded, heterogeneous, complex structure environments or the Brownian motion in an inhomogeneous medium \cite{WEKN:04:B_J, MJD:09:BJ,OFLV:19:FP}. Thus, anomalous diffusion has been paid more attention in recent studies on MC for various potential applications of MCs. MC under anomalous diffusion law in a $1$-dimensional ($1$-D) channel was first considered in \cite{CTJS:15:CL}. 
In \cite{TJS:19:ACCESS}, the authors developed a connectivity model with random time constraints in a $1$-D mobile MC system. The network performance on the error probability in a $2$-D stochastic nanonetwork was investigated in \cite{TJSW:19:COM}. $3$-D concentration-encoded subdiffusive MC and $3$-D subdiffusive MC with an absorbing receiver were considered in \cite{MMM:16:NB} and \cite{HLXGY:20:WCL}, respectively. Additionally, a new mathematical framework for the modeling and analysis of molecular communication under anomalous diffusion was developed in \cite{TJSW:20:COM}. However, most of the studies have focused on analyzing the communication performance of MC with an absorbing receiver, that is, an active reception process is considered.

Normal diffusion-based MC with a \emph{receptor space} (RS) where a receive nanomachine (or a nanosensor) is located inside the RS was considered in $1$-, $2$-, and $3$-D space in \cite{MYCA:14:SP} and was studied with a passive receiver \cite{NCS:14:NB1,DNGNE:17:MBSC}. For more realistic diffusion environments for MC systems, a $3$-D inhomogeneous diffusion medium with non-absorbing receiver, according to Fick’s law with varying diffusivity, has been introduced in \cite{HNWW:19:BioS} and \cite{HW:21:NCN}. These models open a promising and potential research direction in designing and developing nanosensor networks.

In this paper, we consider MC with passive receivers in anomalous diffusion channels. The main contributions of this paper are as follows. We first provide the expected number of observed molecules inside the RS, where the molecules obey anomalous diffusion within arbitrary dimensional spaces. We then derive the peak time of observable molecules. The performance of the MC system is analyzed in terms of bit error rate (BER) in a single-bit interval transmission (SBIT) and a multi-bit interval transmission (MBIT). Finally, we discuss the effects of inter-symbol interference (ISI) on the BER performance by taking into consideration lifetime limited molecules that perish due to enzymes.

\section{Channel and System Models }	\label{sec:2}
\subsection{Channel Model}
In this paper, we consider an isotropic and symmetric $m$-D anomalous diffusion channel based on a space-time fractional diffusion equation \cite[eq.~(13)]{Uch:03:ETP}:\footnote{Note that $\lim_{\beta \rightarrow 1} \frac{t^{-\beta}}{\GF{1-\beta}} =\delta\left(t\right)$, which is corresponding to the initial condition such that $\omega\left(\V{x},0\right)=\delta\left(\V{x}\right)$ \cite{Uch:03:ETP, ZUS:99:ETP}.}
%
\begin{align}	\label{eq:andiff}
	\frac{\partial^\beta}
	{\partial t^\beta}
	\, \omega\left(\V{x},t\right)
	=
	-K
	\,
	\left(-\triangle\right)^{\alpha/2}
	\omega\left(\V{x},t\right)
	+\frac{t^{-\beta}}{\GF{1-\beta}}\delta\left(\V{x}\right)
\end{align} 
where $\omega\left(\V{x},t\right)$ is the fundamental solution (field variable) that represents the probability density function (PDF) of the molecule's position $\V{x} \in \R^m$ at time $t$ where the motion starts at the origin at the initial time;  $K$~[m$^2$/s] is the diffusion coefficient; $\partial^\beta/\partial t^\beta$ is the Riemann–Liouville fractional derivative of order $\beta$  ($0<\beta\leq1$); and $\left(-\triangle\right)^{\alpha/2}$ is the Riesz space fractional derivative of order $\alpha$ ($0<\alpha\leq2$). It is known that the Fourier transform of Riesz space fractional derivative has the form of \cite{ZUS:99:ETP}
\begin{align}
\mathcal{F}\left\{\left(-\triangle\right)^{\alpha/2} f\left(\V{x}\right)\right\}=\sNorm{\V{k}}^\alpha F\left(\V{k}\right), \quad \V{k}\in \R^m
\end{align}
where $\mathcal{F}\left\{f\left(\V{x}\right)\right\}=F\left(\V{k}\right)$. Then, with the initial condition $\omega\left(\V{x},0\right)=\delta\left(\V{x}\right)$ and the boundary condition $\omega\left(\pm\V{\infty},t\right)=0$, the fundamental solution of \eqref{eq:andiff} for $1\leq \alpha\leq 2$ is  \cite{Uch:03:ETP}
\begin{align} \label{eq:fundSol}
\omega\left(\V{x},t\right)
&=
	\frac{1}{\alpha \left(r\sqrt{\pi}\right)^m} 
	\,
	\FoxH{2}{1}{2}{3}{\frac{r}{2K^{\frac{1}{\alpha}}t^{\frac{\beta}{\alpha}}}}{\left(1,\frac{1}{\alpha}\right),\bigl(1,\frac{\beta}{\alpha}\bigr)}{\left(1,\frac{1}{\alpha}\right),\left(\frac{m}{2},\frac{1}{2}\right),\left(1,\frac{1}{2}\right)}
\end{align}
where $r=\sNorm{\V{x}}$ and $H^{\pM,\pN}_{\pP,\pQ}{\left(\cdot\right)}$ denotes the Fox's $H$-function \cite{JSW:15:IT}.\footnote{The $H$-function is a generalization of the Meijer $G$-function. The notation, elementary identities and properties of Fox’s $H$-function can be found in \cite{JSW:15:IT}.}
Note that for normal diffusion $\left(\alpha=2, \beta=1\right)$ in 3-D diffusion, the fundamental solution \eqref{eq:fundSol} can be reduced to \cite{MYCA:14:SP}
\begin{align} 
\omega\left(\V{x},t\right)
	&=
	\frac{1}{2\pi^{3/2}r^3}
	\FoxH{1}{0}{0}{1}{\frac{r}{\sqrt{4Kt}}}{\textrm{--}}{\left(\tfrac{3}{2},\tfrac{1}{2}\right)} \nonumber \\
	&=
	\left(4\pi Kt\right)^{-m/2} \exp\left(-\frac{r^2}{4Kt}\right).
\end{align}
The mean squared displacement of the molecule's movement in the asymptotic limit of large $t$ is proportional to time $t$ as $\bigl<\V{x}\left(t\right)^2\bigr> \propto t^{2\beta/\alpha}$ \cite{TJSW:20:COM}.\footnote{Anomalous diffusion can be classified into various diffusion classes according to the parameters $\alpha$ and $\beta$: i) normal diffusion or standard diffusion ($\alpha=2$, $\beta=1$); ii) quasinormal diffusion ($\alpha=2\beta$); iii) subdiffusion ($2\beta/\alpha <1$); and iv) superdiffusion ($2\beta/\alpha >1$). See \cite[Fig.~2]{TJSW:19:COM} for various types of diffusions.}

\begin{figure}[t!]
\centering
        \includegraphics[width=0.7\textwidth]{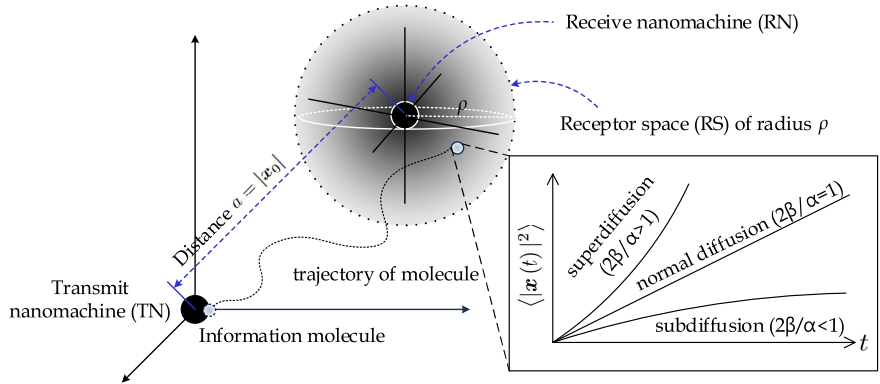}
    \caption{
       An illustration of a 3-D molecular communication system with anomalous diffusion.
    }
    \label{fig:model}
\end{figure}
\subsection{System Model}
We consider an $m$-D MC system including a transmit nanomachine (TN) located at the origin which communicates with a receive nanomachine (RN) located at $\V{x}_0 \in \R^m$, $m\in\left\{1,2,3\right\}$, from TN  by releasing information molecules obey an anomalous diffusion law (see Fig.~\ref{fig:model} for illustrating the 3-D molecular communication system as an example). The transmit information modulates the number of molecules emitted by TN. We consider the burst transmission of an information sequence of finite length. Let $\mathcal{S} \triangleq \left\{s_1,s_2,\ldots,s_{\kappa}\right\}$ be the \emph{information sequence} consisting of $\kappa$ time slots with a bit interval $\Tb$, where $s_i \in\left\{0,1\right\}$  denotes the $i$th transmitted bit, $1\leq i\leq \kappa$. The TN emits $N$ molecules at the beginning of a bit interval to transmit bit `1' whereas no molecule is emitted for bit `0' transmission. 

To decode the information, the RN counts the number of molecules inside the RS of radius $\rho$ at the observation time. Note that the MC channel is assumed to be isotropic and symmetric, we will omit the vector notation of $\V{x}_0$   in all following equations in the rest of this paper. To focus on the characteristics of the anomalous diffusion channel, we assume that the MC system satisfies the following assumptions: i) the number of information molecules for each symbol are controlled by the TN perfectly; ii) the clocks of the TN and RN are synchronized; iii) the information molecules are randomly and freely propagated within the anomalous diffusion medium; and iv) the RN perfectly counts the number of molecules in the RS at the observation time. In addition, to exemplify the anomalous diffusive propagation environment, throughout the remainder of this paper, we set diffusion parameters $\alpha=2$ and $\beta=1$ for normal diffusion, $\alpha=2$ and $\beta=0.5$ for subdiffusion, and $\alpha=1.8$ and $\beta=1$ for superdiffusion with $K=10^{-10}$\,[m$^2$/s], $a=5$\,[$\mu$m], and $\rho=0.5$\,[$\mu$m].
\section{Reception and Detection Processes}
\subsection{Reception Process}
\begin{figure}[t!]
\centering
        \includegraphics[width=0.6\textwidth]{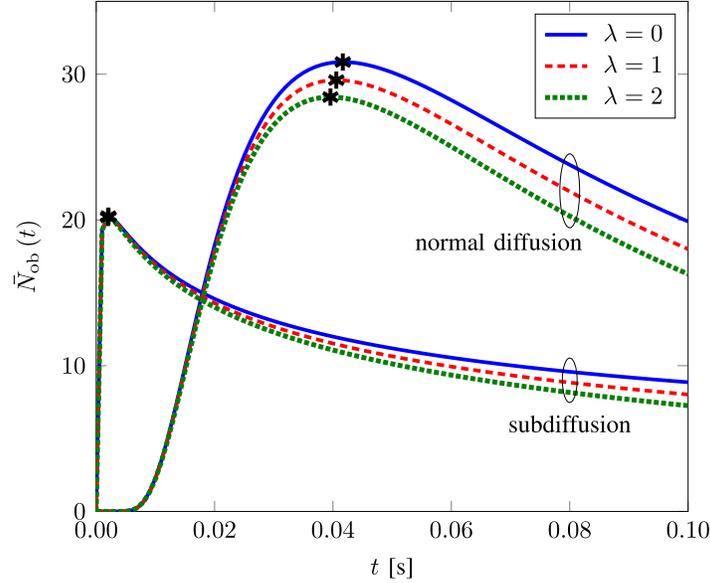}
    \caption{
      Expected number of molecules inside the RS $\bar{N}_{\mathrm{ob}}\left(t\right)$ as a function of $t$ in   normal diffusion and subdiffusion for $\lambda=0,1,2$ when $m=3$ and $N=10^5$.
    }
    \label{fig:Nob:1}
\end{figure}

\begin{figure}[t!]
\centering
        \includegraphics[width=0.6\textwidth]{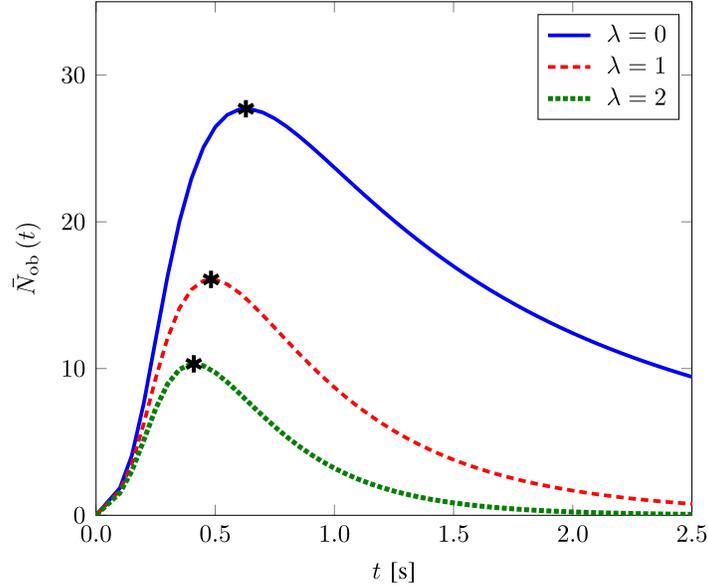}
    \caption{
      Expected number of molecules inside the RS $\bar{N}_{\mathrm{ob}}\left(t\right)$ as a function of $t$ in superdiffusion for $\lambda=0,1,2$ when $m=3$ and $N=10^5$.
    }
    \label{fig:Nob:2}
\end{figure}

For the passive receiver of interest, we assume that the probability that the information molecules are observed inside the RS is equal to the one at the RN located at the center of the RS. In addition, when the TN is sufficiently far from the RS, the precise shape is irrelevant.\footnote{These assumptions are commonly applied for a \emph{passive} receiver model in the context of molecular communication \cite{MYCA:14:SP,NCS:14:NB1,DNGNE:17:MBSC}. For a receiver with an active reception process, the first passage time of molecules plays a key role in determining MC system performance (see \cite{TJSW:19:COM, TJSW:20:COM}).} 
 We also take into consideration lifetime limited molecules whose life expectancy follows an exponential distribution such that\footnote{The life expectancy of molecules can be observed when molecules dissipate due to the presence of enzymes, or due to other chemical reactions in the channel \cite{TJSW:19:COM,NCS:14:NB1,DNGNE:17:MBSC}.} 
\begin{align}
h\left(\tau\right)
&=
\lambda e^{-\lambda \tau}
\end{align}
where $\lambda$ is the degradation rate of the molecules and $\tau$ denotes the molecules' lifetime  \cite{SML:15:WCOM,GAFYLEC:16:WCOM}. It implies that a molecule can be inside the RS at time $t$ only if its lifetime is greater than $t$, that is $\tau \in \left(t,\infty\right)$. Note that $\lambda=0$ corresponds to the no degradation case.
Then, the probability that the information molecules with the degradation rate $\lambda$ stay inside the RS of volume $V_{\rho}$ at the observation time $t$, denoted by $P\left(t | V_{\rho}, \lambda\right)$, is
\begin{align}
P\left(t|V_{\rho},\lambda\right)
&=
	\int_{V_{\rho}}\omega\left(\V{x},t\right)\int_t^{\infty}h\left(\tau\right)d\tau d\V{x} \\
&\approx
	V_{\rho}\omega\left(a,t\right)e^{-\lambda t}
\end{align}
where 
\begin{align}
V_{\rho}
&=
\begin{cases}
2\rho, & \text{ if $m=1$}\\
\pi \rho^2, & \text{ if $m=2$}\\
\frac{4}{3}\pi \rho^3, & \text{ if $m=3$}.
\end{cases}
\end{align}
The expected number of observed molecules inside the RS, denoted by $\bar{N}_{\mathrm{ob}}\left(t\right)$, is given by \cite{NCS:14:NB1,DNGNE:17:MBSC}
\begin{align} \label{eq:Nob}
&\bar{N}_{\mathrm{ob}}\left(t\right) 
=
	V_{\rho}N\omega\left(a,t\right)e^{-\lambda t}.
\end{align}
Since $\bar{N}_{\mathrm{ob}}\left(t\right)$ is the concave function on the interval $t>0$, $\bar{N}_{\mathrm{ob}}\left(t\right)$ obviously has a global maximum value at certain time $t$.\footnote{Since $\omega\left(a,t\right)$ is concave, i.e., $\frac{\partial^2}{\partial^2 t} \omega\left(a,t\right) < 0$ for $t>0$, and $e^{-\lambda t}$ is convex and decreasing, the product of $\omega\left(a,t\right)$ and $e^{-\lambda t}$ is concave. See also Figs.~\ref{fig:Nob:1} and~\ref{fig:Nob:2}).} Let $t_{\mathrm{p}}$ be the \emph{peak time} at $t_\mathrm{p}=\argmax_t \, \bar{N}_{\mathrm{ob}}\left(t\right) $. Then, $t_{\mathrm{p}}$ can be found as 
\begin{align}
	\left.\frac{d \bar{N}_{\mathrm{ob}}\left(t\right) }{d t}\right|_{t=t_{\mathrm{p}}}=0.
\end{align}
\begin{theorem}[Peak Time in $m$-D Anomalous Diffusion] Given the diffusion parameters $\alpha$, $\beta$, $K$, $a$, and $\lambda$, the peak time $t_{\mathrm{p}}$ is the solution of 
\begin{align}	\label{eq:tp}
&\FoxH{1}{3}{4}{3}
{\frac{K^\frac{1}{\beta}t_{\mathrm{p}}}{\left(a/2\right)^{\frac{\alpha}{\beta}}}}
{\left(0,1\right),\bigl(0,\tfrac{1}{\beta}\bigr),\bigl(\tfrac{2-m}{2},\tfrac{\alpha}{2\beta}\bigr),\bigl(0,\frac{\alpha}{2\beta}\bigr)}
{\bigl(0,\tfrac{1}{\beta}\bigr),\left(0,1\right),\left(1,1\right)}
\nonumber \\
&=
	\lambda t_\mathrm{p}
\FoxH{1}{2}{3}{2}
{\frac{K^\frac{1}{\beta}t_{\mathrm{p}}}{\left(a/2\right)^{\frac{\alpha}{\beta}}}}
{\bigl(0,\frac{1}{\beta}\bigr),\bigl(\frac{2-m}{2},\frac{\alpha}{2\beta}\bigr),\bigl(0,\frac{\alpha}{2\beta}\bigr)}
{\bigl(0,\frac{1}{\beta}\bigr),\bigl(0,1\bigr)}.
\end{align}

\begin{proof}
It follows readily from the differential operation of Fox's $H$-function \cite[Property~4]{JSW:15:IT}.
\end{proof}
\end{theorem}
Specifically, we commonly set $\Tb > t_{\mathrm{p}}$ and $t_{\mathrm{p}}$ is equal to $a^2/\left(2mK\right)$ in normal diffusion when $\lambda=0$ \cite{MYCA:14:SP, NCS:14:NB1}.\footnote{It is noteworthy that, within a given anomalous diffusion channel, the peak time $t_\mathrm{p}$ highly depends on the distance $a$ and the diffusion coefficient $K$.}

The expected number of molecules inside the RS $\bar{N}_{\mathrm{ob}}\left(t\right)$ is shown for normal diffusion and subdiffusion in Fig.~\ref{fig:Nob:1} and for superdiffusion in Fig.~\ref{fig:Nob:2} as a function of $t$ for $\lambda=0$, $1$, and $2$ when $m=3$ and $N=10^5$. We can see that $\bar{N}_{\mathrm{ob}}\left(t\right)$ increases over time until reaching its peak time $t_\mathrm{p}$, and then monotonically decreasing with time. For example, $t_{\mathrm{p}}$ for normal diffusion, subdiffusion, and superdiffusion are equal to $0.0417$, $0.0021$ and $0.6287$ when $\lambda=0$. Since the subdiffusion information molecules are most rapidly propagated in space while the superdiffusion molecules are spreading slowly in space, the largest $t_\mathrm{p}$ is observed in superdiffusion while subdiffusion has the smallest $t_{\mathrm{p}}$. Note that with sufficiently large $a$ and $K$, superdiffusion is expected to have the largest $t_{\mathrm{p}}$ \cite{TJS:19:ACCESS}. Moreover, we can observe that $\bar{N}_{\mathrm{ob}}\left(t\right)$ decreases with the degradation rate $\lambda$, for example, $t_\mathrm{p}$ for $\lambda=1$ and $2$ in Fig.~\ref{fig:Nob:2} are 0.4824 and 0.4099, respectively.
\subsection{Detection Process}
Let $t_{\mathrm{o}}$ be the observation time that the RN counts the number of information molecules inside the RS after the beginning of the $i$th bit interval in order to decode the bit $s_i$, $1\leq i\leq \kappa$, $t_{\mathrm{o}} \in \left(0,\Tb\right]$. Let $\mathcal{S}_{i}  \triangleq \left\{s_{1},s_{2},\ldots,s_{i-1}\right\}$ be an \emph{ISI sequence} of the $i$th bit transmission consisting of all previous bit transmission $s_j$, $1\leq j<i$. The probability that the molecule emitted at the $j$th bit transmission inside the RS at $t_\mathrm{o}$ is equal to $P\left(\left(i-j\right)\Tb+t_{\mathrm{o}}|V_{\rho},\lambda\right)$. Then, the total number of molecules counted by the RN for the detection of the $i$th bit transmission, denoted by $y_{i}$, is distributed as 
%
\begin{align} \label{eq:yi}
y_{i}
&\sim
	\sum_{j=1}^i s_j\Binom{N}{P\left(\left(i-j\right)\Tb+t_{\mathrm{o}}|V_{\rho},\lambda\right)}
\end{align}
where $\Binom{x}{y}$ denotes the binomial distribution with mean $xy$ and variance $xy\left(1-y\right)$.  
Since the sum of binomial distribution is mathematically not tractable, it can be alternatively approximated by either the Gaussian distribution or Poisson distribution. It has been shown that the Poisson approximation was indistinguishable from the binomial distribution, whereas a notable loss in accuracy was seen in the Gaussian approximation \cite{NCS:14:NB1}. Hence, we consider a Poisson approximation for the sum of binomial random variables. Then, we have 
\begin{align} \label{eq:appPoi}
y_{i}\sim \mathrm{Pois}\Biggl(\sum_{j=1}^i s_j NP\left(\left(i-j\right)\Tb+t_{\mathrm{o}}|V_{\rho},\lambda\right)\Biggr)
\end{align}
where $\PV{x}$ denotes the Poisson distribution with the mean $x$.
The $i$th information bit $s_i$ can be decoded by using the following detection rule:
\begin{align}
\hat{s}_i
&=
\begin{cases}
1, \quad \text{if $y_{i}\geq \gamma_i$} \\ 
0, \quad \text{if $y_{i} < \gamma_i$}
\end{cases}
\end{align}
where $\gamma_i > 0$ is the decision threshold for the $i$th transmitted bit. 
\section{Error Probability}
\subsection{Single-Bit Interval Transmission} \label{sec:SBIT}
We first consider a SBIT with $\kappa=1$. 
In this case, the total number of molecules counted by the RN at $t_\mathrm{o}$ for the detection of the $i$th bit transmission is distributed as
\begin{align}
y_{i}
&\sim
	\Binom{N}{P\left(t_{\mathrm{o}}|V_{\rho},\lambda\right)}.
\end{align}

\begin{theorem}[BER of SBIT]\label{thr:Pe:Bi}
For equally-likely information bits $s_i$, the BER $\Pe$ of SBIT with the decision threshold $\gamma_i$ is given by
\begin{align} \label{eq:Pe:Bi}
\Pe&= \frac{1}{2}I_{q}\left(N-\gamma_i+1,\gamma_i\right)
\end{align}
where $q=1-P\left(t_{\mathrm{o}}|V_{\rho},\lambda\right)$, and $I_x\left(\cdot,\cdot\right)$ is the regularized incomplete beta function \cite[eq. (8.392)]{GR:07:Book}.
\begin{proof} Since no molecule is emitted when bit `0' is transmitted, with $\gamma_i >0$, $P_e$ is given by
\begin{align}
\Pe&=\frac{1}{2}\Prob{\hat{s}=0|s=1} =
      \frac{1}{2}\Prob{y_{i}<\gamma_i|s=1}.
\end{align}
From which and the cumulative distribution function (CDF) expression of binomial distribution, we arrive at the desired result.
\end{proof}
\end{theorem}

\begin{remark} [Optimal Observation Time] \label{cor:to}Let $t^\star_{\mathrm{o}}$ be the optimal observation time in which $t^\star_\mathrm{o}=\argmin_t  \Pe$.\footnote{Assume that $V_{\rho} \ll 1$. Then, $1-V_{\rho}\omega\left(a,t\right) e^{-\lambda t}$ be convex for $t>0$.} Then, $t^\star_{\mathrm{o}}$ is the solution of
\begin{align}
	\left.\frac{d \left(1-V_{\rho}\omega\left(a,t\right)e^{-\lambda t}\right) }{d t}\right|_{t=t^\star_{\mathrm{o}}}=0.
\end{align}
Then, it is readily shown that $t^\star_{\mathrm{o}}=t_{\mathrm{p}}$.
\end{remark}
\begin{remark}[Optimal Decision Threshold] Since $\Pe$ is the CDF of binomial distribution, the optimal decision threshold that minimizes $\Pe$ is obviously equal to $1$. With $\gamma_i=1$, $\Pe$ in~\eqref{eq:Pe:Bi} reduces to
$
\Pe=\frac{1}{2} \left(1-P\left(t_{\mathrm{o}}|V_{\rho},\lambda\right)\right)^N
$.
\end{remark}

\begin{remark}[Transmit Diversity Gain]
Let 
\begin{align}   \label{def:hm:slope}
    \xi
    \triangleq
    \lim_{N\rightarrow \infty}
    \frac{-\log \Pe}{N}
\end{align}
be the \emph{transmit diversity gain} of the BER $\Pe$.  Then by applying L'H\^{o}pitals's rule for $0<p<1$, and with the fact that
$$
    \lim_{x\rightarrow \infty}\frac{-\log \RIBF{p}{x}{y} }{x}=-\log p
$$
we have
\begin{align}
\xi
&=
	\log\left(\frac{1}{1-P\left(t_{\mathrm{o}}|V_{\rho},\lambda\right)}\right).
\end{align}
This reveals that the reliability of the MC system can be enhanced from the consumption of more molecular resources, which is synonymous with the transmit diversity gain in multi-antenna wireless communication systems. It is also noteworthy that $\xi$ is not a function of the detection threshold.
\end{remark}
\subsection{Multi-Bit Interval Transmission}
In a MBIT with $\kappa>1$, we assume that the RN can keep the transmitted sequence history $\mathcal{S}_{i}$ (the memory detector) \cite{HLXGY:20:WCL}.
\begin{theorem}[Conditional BER of MBIT] Let 
\begin{align}
\mu^{\left(k\right)}_{i} &= \sum_{j=1}^{i-1+k} s_j NP\left(\left(i-j\right)\Tb+t_{\mathrm{o}}|V_{\rho}
\lambda\right)
\end{align}
be the average of $y_{i}$ when the information bit $s_i = k$, $k \in\left\{0,1\right\}$. For equally-likely information bits $s_i$, the conditional BER of the $i$th bit in the burst transmission given $\mathcal{S}_{i}$, denoted by $\PeSL$, with the decision threshold $\gamma_i$ is given by\footnote{$\PeSL$ is practically important for a realistic detector based on the detection result of $\mathcal{S}_i$ since it can be used to evaluate the error rate of the burst.} 
\begin{align}  
\PeSL
&=
\frac{1}{2}
\left(1-
\frac{
\sum_{k=0}^{1}
\left(-1\right)^k
\GF{\floor{\gamma_i},\mu^{\left(k\right)}_{i}}
}{\floor{\gamma_i-1}!}
\right)
\end{align}
where $\GF{\mu,\sigma}$ is the upper incomplete Gamma function \cite[eq.~(8.350)]{GR:07:Book} 
and $\floor{x}$ denotes the floor function.
\begin{proof}  For equiprobable bits `0' and `1', we have
\begin{align*}
\PeSL &=
\frac{1}{2}\left(\Prob{y_{i}\geq\gamma_i|s_i=0,\mathcal{S}_{i}}
+
\Prob{y_{i}<\gamma_i|s_i=1,\mathcal{S}_{i}}\right).
\end{align*}
From which and the CDF expression of Poisson distribution, we obtain the desired result.
\end{proof}
\end{theorem}
Note that it is needed to average on all possible ISI sequences with their probabilities of occurrence to calculate the average BER. We only consider $\mathcal{S}_i=\B{1}_{i-1}$ in the following section for the most serious ISI performance analysis where $\B{1}_n$ denotes an all-one sequence of $n$ elements.
\begin{remark} For the memory detector, the optimal $\gamma_i$ can be found using the maximum likelihood such that \cite{HLXGY:20:WCL}
\begin{align} \label{eq:thr:ML}
\gamma_i=\frac{\mu^{\left(0\right)}_{i} -\mu^{\left(1\right)}_{i} }{\log{\mu^{\left(0\right)}_{i} }-\log{\mu^{\left(1\right)}_{i}}}.
\end{align}
\end{remark}
%


\begin{figure}[t!]
\centering
        \includegraphics[width=0.6\textwidth]{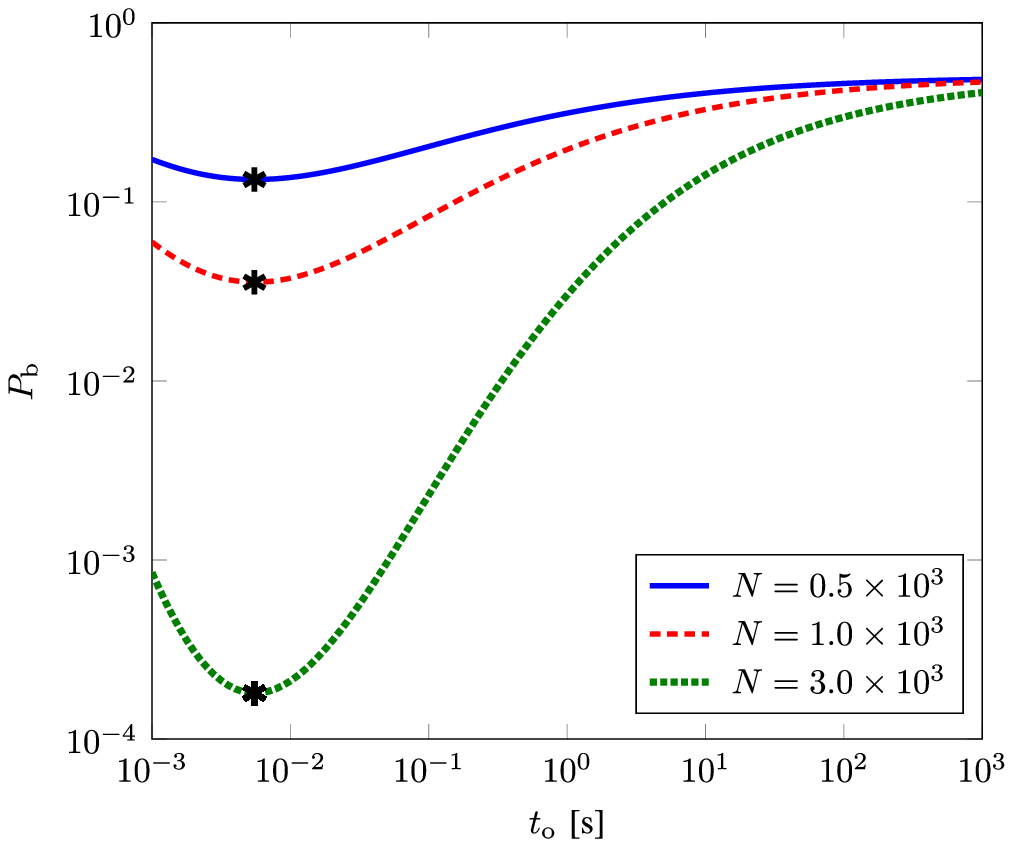}
    \caption{
    BER $\Pe$ of SBIT as a function of $t_{\mathrm{o}}$ in subdiffusion when $m=3$.
    }
    \label{fig:Pb:ts}
\end{figure}

\begin{figure}
\centering
        \includegraphics[width=0.62\textwidth]{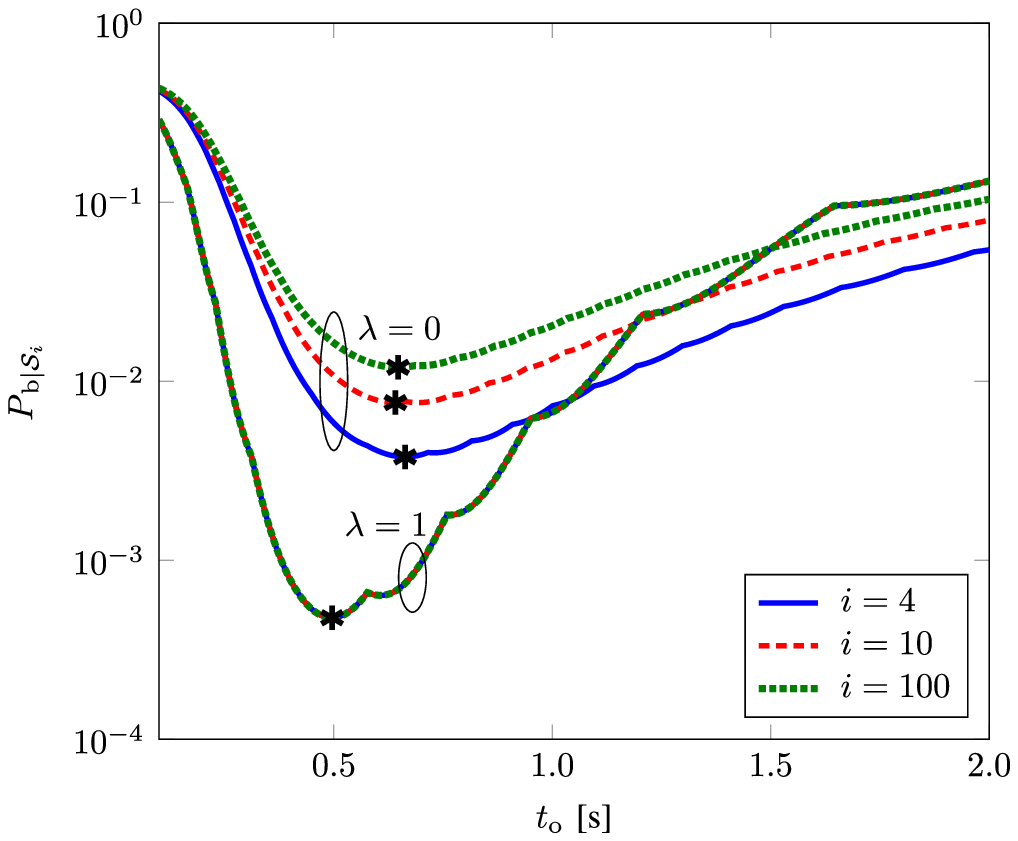}   
    \caption{
       BER $\PeSL$ of $s_i$ in MBIT as a function of $t_\mathrm{o}$ in superdiffusion when $m=3$.
    }
    \label{fig:Peto}
\end{figure}
%
\subsection{Numerical Examples}
Fig.~\ref{fig:Pb:ts} shows the BER $\Pe$ of SBIT as a function of observation time $t_{\mathrm{o}}$ in subdiffusion for i) $N=0.5\times 10^3$, ii) $1.0\times 10^3$, and iii) $3.0 \times 10^3$ when $\lambda=0$, $\gamma_i=1$, and $m=2$. We observe that the BER decreases as the number of emitted molecules $N$ increases. In addition, the minimum $\Pe$ is achieved at the optimal observation time $t_{\mathrm{o}}=t_{\mathrm{p}}=0.0055$\,[s] as described in Remark~1 and Remark~2. 
Fig~\ref{fig:Peto} shows the BER $\PeSL$ of $s_i$ in MBIT with $\mathcal{S}_{i} = \B{1}_{i-1}$ as a function of $t_\mathrm{o}$ in superdiffusion for i) $i=4$, ii) $i=10$, and iii) $i=100$ with $\lambda=0$ and $\lambda=1$ when $\Tb=2$, $m=3$, and $N=10^5$. The decision threshold $\gamma_i$ is found using \eqref{eq:thr:ML}. We can see that the observation time that minimizes $\PeSL$ is not equal to the one given in Remark~1 due to the ISI effect. Indeed, the large size of the interference sequence gives BER performance degradation while the lifetime limited interfering molecules reduces the ISI effect on the BER performance. It can be expected that if $t_{\mathrm{o}}$ is near to $\Tb$, though the less ISI is cumulated, but the number of observed information molecules decreases as well. Therefore, it leads to a higher BER in general. Furthermore, it is seen that the BER converges for $i$ $\geq 4$ where $\lambda=1$. However, the performance consistently degrades as $i$ increases when $\lambda =0.$ It means that a reliable continuous transmission can be achieved only with lifetime limited molecules.

\begin{figure}[t!]
\centering
        ~~\includegraphics[width=0.62\textwidth]{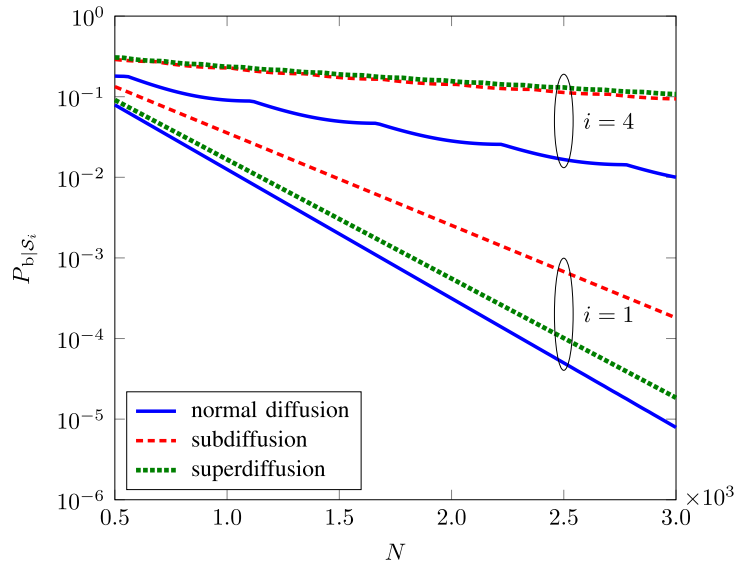}
    \caption{
       BER $\PeSL$ of $s_i$ in SBIT and MBIT as a function of $N$ when $m=2$. 
       }
    \label{fig:Pb:N}
\end{figure}

The BER $\PeSL$ of $s_i$ in SBIT and MBIT ($i=4$ with $\mathcal{S}_{i} = \B{1}_{i-1}$) as a function of $N$ is shown in Fig.~\ref{fig:Pb:N} for i) normal diffusion, ii) subdiffusion, and iii) superdiffusion when $\lambda=0$, $t_{\mathrm{o}}=t_{\mathrm{p}}$, $m=2$. For SBIT, we set $\gamma_i=1$, while $\gamma_i$ can be found using \eqref{eq:thr:ML} for MBIT. It can obviously be seen that the BER decreases as $N$ increases. In the case of SBIT, the transmit diversity gain $\xi$ are equal to $0.00160$, $0.00115$, and $0.00148$ for normal diffusion, subdiffusion, and superdiffusion, respectively, as stated in Remark~3. It reveals that the BER in normal diffusion outperforms that in other diffusion scenarios for the large $N$ in this example. The transmit diversity gain for MBIT, which can be evaluated numerically, is tabulated in Table~\ref{table:1} for $i=2,3,4$ with $\mathcal{S}_{i}= \B{1}_{i-1}$.  It is shown that the ISI decreases transmit diversity gain significantly, as expected.

\begin{figure}[t!]
\centering
        \includegraphics[width=0.6\textwidth]{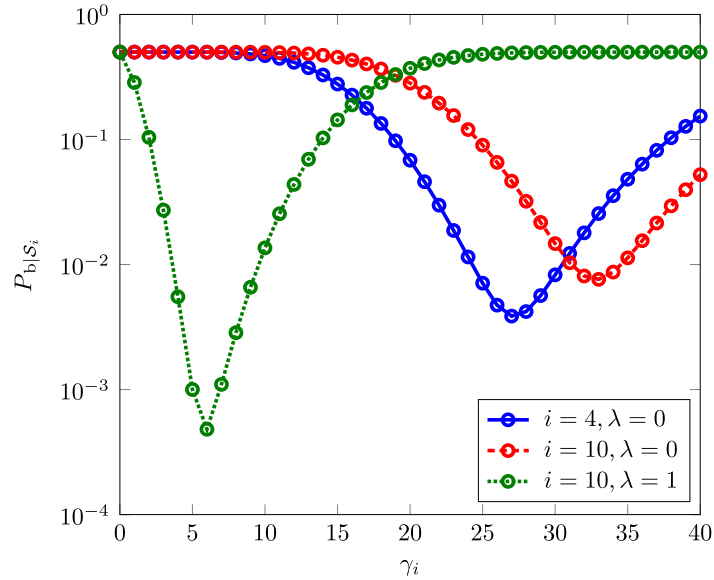}   
    \caption{
       BER $\PeSL$ of $s_i$ in MBIT as a function of $\gamma_i$ in superdiffusion when $m=3$.
    }
    \label{fig:PeL}
\end{figure}

\begin{table}[t!]
\caption{Transmit Diversity Gain $\xi$} \centering
\label{table:1}
\begin{tabular}{lcccc}
\toprule
& \multicolumn{4}{c}{$i$ } \\
 \cline{2-5} \\[-0.5cm]
Diffusion scenario            & $1$ & $2$ & $3$ & $4$    \\[-0.05cm]
\midrule
Normal diffusion    & $0.00160$ & $0.00057$ & $0.00051$ & $0.00050$   \\
Subdiffusion        & $0.00115$ & $0.00030$ & $0.00023$ & $0.00019$ \\
Superdiffusion      & $0.00148$ & $0.00026$ & $0.00021$ & $0.00018$  \\
\bottomrule
\end{tabular}
\end{table}
The  BER $\PeSL$ of $s_i$ in MBIT with $\mathcal{S}_{i} = \B{1}_{i-1}$ is shown in Fig.~\ref{fig:PeL} as a function of $\gamma_i$ for i) $\lambda=0$ and ii) $\lambda=1$ in superdiffusion when $\Tb=2$, $t_{\mathrm{o}}=t_{\mathrm{p}}$, $m=3$, and $N=10^5$. With the observation time $t_{\mathrm{o}}=t_{\mathrm{p}}$, the decision threshold that minimizes $\PeSL$ is varying with respect to the number of expected interfering molecules. For example, with $\lambda=0$, $\gamma_i$ is equal to $27$ for $i=4$, and $\gamma_i$ is equal to $33$ for $i=10$. It can be found the optimal decision threshold is numerically equal to the ceiling of $\gamma_i$ given in Remark~4. We also find it noteworthy that the ISI effects can be controlled simply by taking the lifetime limited molecules into consideration.
\section{Conclusion}
In this paper, we studied the MC system in anomalous diffusion channels with the RS at the receiver side. The peak time of the expected number of observed molecules inside the RS has been presented. In addition, we analyzed the performance of the MC system in terms of BER. It has been shown that the peak time is the optimal observation time that minimizes the BER in the SBIT. In the MBIT, the BER has been derived with a given ISI sequence. The corresponding observation time and decision threshold that are minimizing BER can be found numerically. It has been shown that introducing lifetime limited molecules is an effective approach for the alleviation of the ISI in the MBIT. A more realistic diffusion environment for various applications of MC systems, i.e., anomalous diffusion in an inhomogeneous medium, will be developed in future studies.

\bibliographystyle{IEEEtran}

\end{document}